\documentclass[useAMS,usenatbib]{mn2e}
\usepackage{graphicx}
\usepackage{latexsym}
\usepackage{subfigure}
\usepackage{url}
\usepackage{psfig}

\title{Ionizing feedback from massive stars in massive clusters: Fake bubbles and untriggered star formation}

\author[James E. Dale, Ian Bonnell]{James E. Dale$^{1,2}$\thanks{E-mail: jim@ig.cas.cz (JED)}, Ian Bonnell$^{3}$\\
$^{1}$Astronomical Institute, Academy of Sciences of the Czech Republic, Bocni II 1401/2a, 141 31 Praha 4, Czech Republic\\
$^{2}$Excellence Cluster `Universe', Boltzmannstr. 2, 85748 Garching, Germany\\
$^{3}$School of Physics and Astronomy, Univerity of St. Andrews, North Haugh, St. Andrews, UK}

\begin{document}

\pagerange{\pageref{firstpage}--\pageref{lastpage}} \pubyear{2006}

\maketitle

\label{firstpage}

\def\mnras{MNRAS}
\def\apj{ApJ}
\def\aap{A\&A}
\def\apjl{ApJL}
\def\apjs{ApJS}
\def\bain{BAIN}
\def\araa{ARA\&A}
\def\pasp{PASP}
\def\aj{AJ}
\def\pasj{PASJ}
\def\ga{\sim}

\begin{abstract}
We use Smoothed Particle Hydrodynamics to simulate the formation of a massive ($10^{6}$M$_{\odot}$) stellar cluster system formed from the gravitational collapse of a turbulent molecular cloud. We investigate the hierarchical clustering properties of our model system and we study the influence of the photoionizing radiation produced by the system's multiple O--type stars on the evolution of the protocluster. We find that dense gas near the ionizing sources prevents the radiation from eroding the filaments in which most of the star formation occurs and that instead, ionized gas fills pre--existing voids and bubbles originally created by the turbulent velocity field.
\end{abstract}


\section{Introduction}
Massive clusters and OB associations associated with giant HII regions such as R136 in 30 Doradus \citep[e.g.][]{1999A&A...347..532S,2006AJ....131.2140T,2009ApJ...694...84I}, the central clusters in NGC 3603 \citep[e.g.][]{2001RMxAA..37...39T,2006AJ....132..253S} and NGC 604 \citep[e.g][]{2008ApJ...685..919T}, and Trumpler 14, 15 and 16 and Collinder 228 \citep[e.g.][]{2003MNRAS.339...44T,2006MNRAS.367..513T} in the Carina HII region, each containing tens to hundreds of O--type stars, are ideal places to test our understanding of star formation and the effects of feedback from the high--mass end of the IMF. Star formation in such regions appears to be hierarchical, with structure extant over a wide range of length and mass scales. The dynamics of these regions, or at least of their remaining gas content, seem to be dominated by feedback from O--stars in the form of giant or multiple HII regions, wind--blown bubbles and supernova shells. These phenomena drive powerful outflows, expelling mass from the embedded clusters and perhaps unbinding them. Several authors \citep[e.g][]{2003ARA&A..41...57L,2005ApJ...631L.133F} have concluded that the majority of star clusters do not survive their embedded phase and the loss of large quantities of gas via outflows offers an attractive explanation for this phenomenon.\\
\indent The dispersal of clouds by the action of feedback has been studied by many authors. \cite{1980ApJ...235..986H} and later \cite{1997MNRAS.284..785G}, \cite{2003MNRAS.338..673B}, \cite{2003MNRAS.338..665B} and \cite{2006MNRAS.373..752G} considered the dynamical effect on clusters of removing gas left over by the star formation process, while, e.g., \cite{1987sbge.proc..467L}, \cite{1988ARA&A..26..145T}, \cite{1994ApJ...436..795F} and \cite{2002ApJ...566..302M} consider the self--limiting effect of feedback on the star--formation efficiency in clouds. These authors did not, however,  model the process of gas removal itself. \cite{2001MNRAS.323..988G} performed hybrid N--body/SPH calculations in which they modelled the interaction of the gas ad stars self--consistently, but the expulsion of the gas, by heating or by driving a shockwave through it, were still parameterized. It is clear that an understanding of the process by which stellar systems lose gas is crucial.\\
\indent \cite{1979A&A....71...59T} introduced the concept to the champagne flow, in which an O--star born near the edge of a cloud creates a cavity around itself through which gas eroded from the wall of the cloud flows at high velocity into the ISM. Using an analytic model, \cite{1979MNRAS.186...59W} found that this process would result in star formation efficiencies at the scale of whole giant molecular clouds would be $\sim5\%$. 2--D numerical simulations by \cite{1979ApJ...233...85B} also showed that champagne flows could be an efficient dispersal mechanism and they clearly do play such a role in many systems, such as Orion \citep{2005ApJ...627..813H}. However, \cite{1980A&A....90...65M} and \cite{1989A&A...216..207Y} found that the disruption of clouds by HII regions is strongly impeded or prevented if the clouds are in freefall and if the ionizing sources are located near the centres of the clouds. As noted above, observations often indicate that massive stars are more likely to be found near the centres of star--forming regions, or at any rate at the centres of subclusters, and many simulations reproduce this result. This can be the result of competitive accretion \citep[e.g][]{2003MNRAS.343..413B} or rapid mass segregation \citep[e.g][]{2007ApJ...655L..45M}.\\
\indent Additionally, the above--cited models of champagne flows treat the gas as being initially uniform or at least smoothly--varying. In contrast, simulations and observations \citep[e.g.][]{2007MNRAS.380.1497L} reveal that massive stars are often embedded in highly inhomogeneous gas from which they are continually accreting and which can severely limit the effects of feedback \citep[e.g][]{2010ApJ...719..831P}. The complex structure of the gas may derive from turbulence \cite[e.g][]{2001ApJ...556..837K}, or from gravitational amplification of even very small quantities of noise in the density field \citep{2002MNRAS.336..659B}. The effects of feedback on such systems demands the use of 3--D self--gravitating simulations such as those carried out by, e.g, \cite{2005MNRAS.358..291D} or \cite{2010ApJ...719..831P}. \cite{2005MNRAS.358..291D} found that strongly inhomogenous and anisotropic accretion flows onto an ionizing source strongly limit the effect of ionizing feedback, with radiation only being able to escape the immediate vicinity of the source over a limited fraction of the sky, in strong contrast to 1--D models of HII regions as Str\"omgren spheres.\\
\indent Feedback is also thought to be responsible for triggering additional star formation in many regions and there is much literature on this topic \citep[e.g][]{2003A&A...408L..25D,2006A&A...446..171Z,2008ApJ...688.1142K,2009A&A...503..107P}. However, identifying triggered star formation from an observational perspective is extremely difficult and morphological indicators, such as shell--like structures of stars and gas, or the close proximity of young stellar objects to HII regions must often be relied upon.\\
\indent In this paper, we investigate the evolution of a massive cluster (or, more accurately, a massive complex of clusters) formed from the gravitational fragmentation of a 10$^{6}$M$_{\odot}$ turbulent molecular cloud. We examine the clustering properties of the stellar systems formed in the cloud and we include the effects of the photoionizing radiation from the multiple O--type stars hosted by the clusters on the highly inhomogeneous gas remaining in the system as a first step in understanding the transition between large--scale embedded star formation and open clusters or OB associations. Our numerical methods are presented in Section 2. Section 3 contains our results, which are then discussed in Section 4. Section 5 contains our conclusions.\\
\section{Numerical methods}
We use the well--known Smoothed Particle Hydrodynamics technique \citep{1992ARA&A..30..543M}, which is ideal for studying the evolution of molecular clouds and embedded clusters. We use a version of the Benz code \citep{1990nmns.work..269B} and the standard artificial viscosity prescription, with $\alpha=1$, $\beta=2$. Particles are evolved on individual timesteps. The code is a hybrid N--body SPH code in which stars are represented by point--mass sink particles \citep{1995MNRAS.277..362B}. Self--gravitational forces between gas particles are calculated using a binary tree, whereas gravitational forces involving sink--particles are computed by direct summation. Sink particles are formed dynamically and may accrete gas particles and grow in mass.\\
\indent The thermodynamics of the gas are treated using a modified Larson--type piecewise equation of state \citep{2005MNRAS.359..211L} comprised of three barotropic equations of state
\begin{equation}
P = k \rho^{\gamma}
\end{equation}
where
\begin{equation}
\begin{array}{rlrl}
\gamma  &=  0.75  ; & \hfill &\rho \le \rho_1 \\
\gamma  &=  1.0  ; & \rho_1 \le & \rho  \le \rho_2 \\
\gamma  &=  1.4  ; & \hfill \rho_2 \le &\rho \le \rho_3 \\
\gamma  &=  1.0  ; & \hfill &\rho \ge \rho_3, \\
\end{array}
\end{equation}
and $\rho_1= 5.5 \times 10^{-19} {\rm g\ cm}^{-3} , \rho_2=5.5 \times10^{-15} {\rm g\ cm}^{-3} , \rho_3=2 \times 10^{-13} {\rm g\ cm}^{-3}$. The low--density cooling part of the equation of state mimics the effects  
of line cooling. The $\gamma=1.0$ segment approximates the effect of dust cooling whilst the $\gamma=1.4$ segment mimics the effects of when collapsing cores become optically thick to infrared radiation. The final isothermal phase of the equation of state is simply in order to allow sink-particle formation to occur, which requires a subvirial collapsing fragment.\\
\indent In these simulations, the model cloud has a mass of $10^{6}$M$_{\odot}$ and is seeded with a Kolmogorov turbulent velocity field whose total kinetic energy is equal in magnitude to the cloud's initial gravitational binding energy, so that the cloud is marginally bound and is therefore expected to form stars rapidly and efficiently in the absence of feedback. The initial mean Jeans mass is $\sim260$ M$_{\odot}$, so that the initial conditions contain $\sim4000$ thermal Jeans masses and we would therefore expect of order this number of fragments to form if the system were isothermal. However, the mean intitial temperature is $\sim40$K, and the collapsing regions cool to 10K, and the Jeans mass in the dense filaments that form is therefore lower.\\
\indent We model the cloud at two resolutions, using 2.5$\times$10$^{7}$ and $10^{6}$ particles. We use the higher resolution calculation to examine the clustering properties of the star formation in the system, and the low--resolution run to study the effects of ionization. Due to the large number of ionizing sources, we found that simulating ionization  in the high--resolution run was too time--consuming. Mass resolution in the high--resolution calculation was 4M$_{\odot}$ \citep{1997MNRAS.288.1060B}, but in the low--resolution simulation is rather large at $100$M$_{\odot}$. The sink particles formed in this simulation cannot therefore be regarded as stars, but are instead small clusters. If such a cluster is sufficiently massive, it will host one or more O--stars. For all sink particles whose mass exceeds 600 M$_{\odot}$ we compute the number N$_{30}$ of $30$ M$_{\odot}$ O--stars assuming that the clusters have a Salpeter IMF with minimum and maximum stellar masses of $0.1$M$_{\odot}$ and $100$M$_{\odot}$. We then take the ionizing luminosity of each cluster to be $N_{30}\times10^{49}$ ionizing photons per second.\\
\indent The effects of ionization on the gas are simulated using a modified version of the Str\"omgren volume algorithm described in \cite{2007MNRAS.382.1759D} to identify ionized particles, whose temperatures are then set to 10$^{4}$K and treated isothermally. The algorithm has been modified in two ways. Firstly, the original algorithm updates the ionization fraction of all particles on the shortest dynamical timestep in use by any particle in the code. Instead, the new code only updates the ionization states of particles when they are being dynamically updated, so that the ionization timestep of each particle is taken to be equal to its dynamical timestep.\\
\indent Secondly, it was necessary to modify the ionization algorithm to allow simulation of the effects of multiple ionizing sources. In the case that no two HII regions overlap, no changes to the algorithm would be required, but if the ionized regions excited by two or more different sources overlap, a given particle may be receiving flux from more than one source. For an ionized particle, the recombination rate per unit volume of the ionized gas making up the particle is $\alpha_{\rm B}n^{2}$, where $n$ is the number density of ions and $\alpha_{\rm B}$ is the case--B recombination coefficient. To keep the particle ionized, ionizing photons must be subtracted from radiation beams passing through the particle. The problem that must be solved is to determine how many photons should be subtracted from each beam. We solve this problem in a crude way by assuming that all sources illuminating a given ionized particle are equally responsible for keeping it ionized. Strictly, the global ionization state of the whole cloud should then be found by iteration, but this is time--consuming. Instead, at a given timestep, if a gas particle is receiving radiation from $N$ ionizing sources, when the particle's ionization state is next updated, we assume that the volume recombination rate subtracted from the radiation beam of each source passing through the particle is $\alpha_{\rm B} n^{2}/N$. We therefore assume that the recombination load is shared equally among all the sources illuminating a given particle.\\
\section{Results}
\subsection{Clustering}
\begin{figure}
\includegraphics[width=0.45\textwidth]{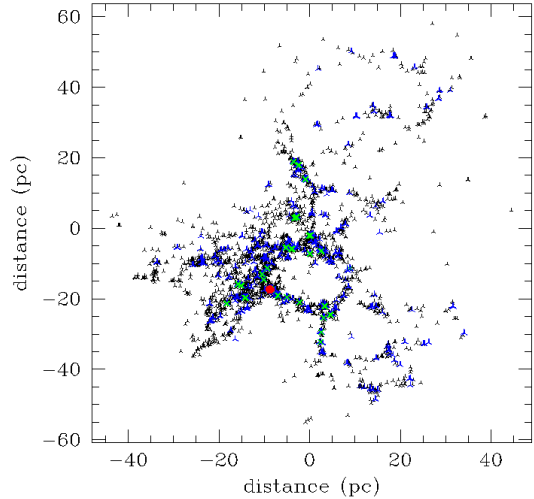}
\caption{Locations of sink particles in the high--resolution calculation. Coloured symbols indicate the centres of clusters whose densities exceed $10^{3}$ (blue), $10^{4}$ (green) and $10^{5}$ (red) M$_{\odot}$ pc$^{-3}$.
\label{fig:OBclus}}
\end{figure}

\begin{figure}
\includegraphics[width=0.45\textwidth]{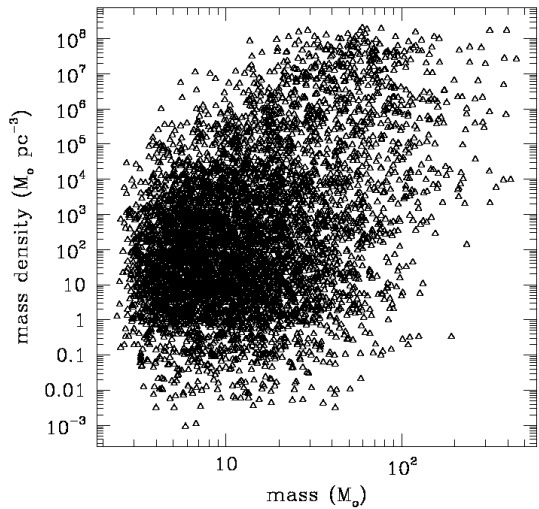}
\caption{Plot of cluster density (as computed by the mass within 1pc) against cluster mass. \label{fig:clus_1pc}}
\end{figure}

\begin{figure}
\includegraphics[width=0.45\textwidth]{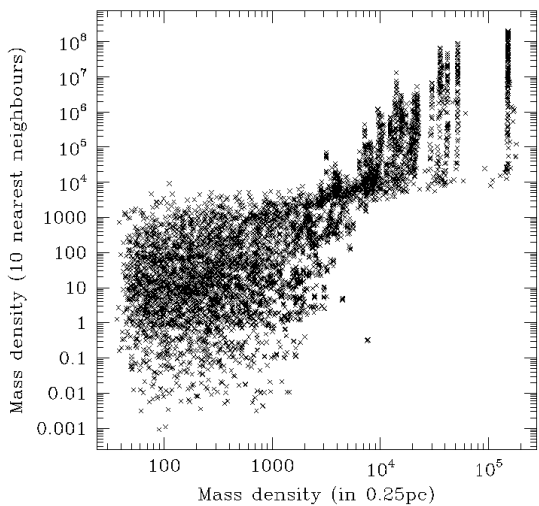}
\caption{Plot of cluster density (as computed by finding the volume containing 10 stars) against cluster mass. \label{fig:clus_10neigh}}
\end{figure}

Both the low resolution and the high resolution clouds evolve in a similar fashion over the 4.3 Myr that the evolutions were followed (without feedback). The turbulence in the gas generates significant internal structure in the  
form of filaments. Regions within the filamentary structure become self-gravitating and collapse to be replaced by sink particles (which represent stars in the high--resolution simulation, and small clusters in the low--resolution run).  
Larger-scale regions become gravitationally bound due to the shock dissipation of kinetic energy, allowing clusters or groups of clusters to form at the junctions of the filaments. These forming clusters continue to grow by accreting infalling gas and other sink particles. The clustering properties of the model are of intrinsic interest for comparison to real massive star forming regions, but also since the distribution of clusters and massive stars determines to a large extent the influence of feedback.\\
\indent The low resolution run created  601 sink-particles whereas the high-resolution simulation created 5750 sinks. In Figure \ref{fig:OBclus}, we plot the location of all stars at a time of 4.3Myr in this simulation. Amongst the sinks are several very significant clusterings including the most massive cluster with a mass density in excess of $10^5$ M$_{\odot}$  pc$^{-3}$ inside 0.25 pc. Figures \ref{fig:clus_1pc} and \ref{fig:clus_10neigh} show the mass density for each sink in the high--resolution computation computed from the mass within 1pc of that sink, and from the distance to its tenth nearest neighbouring sink respectively. We see from these figures that most of the more massive sinks are in highly clustered regions with local mass densities reaching $10^8$ M$_{\odot}$ pc$^{-3}$. In addition to the high clustering of most of the massive sinks, we see a number of relatively massive sinks with masses $\ge 100$ M$_{\odot}$ that are in low density regions as measured from the distance to their tenth nearest neighbour. Such sinks likely represent small clusters that may contain what appear to be isolated massive stars in the context of the whole system.\\
\subsection{Effects of ionization feedback}
\begin{figure}
\includegraphics[width=0.45\textwidth]{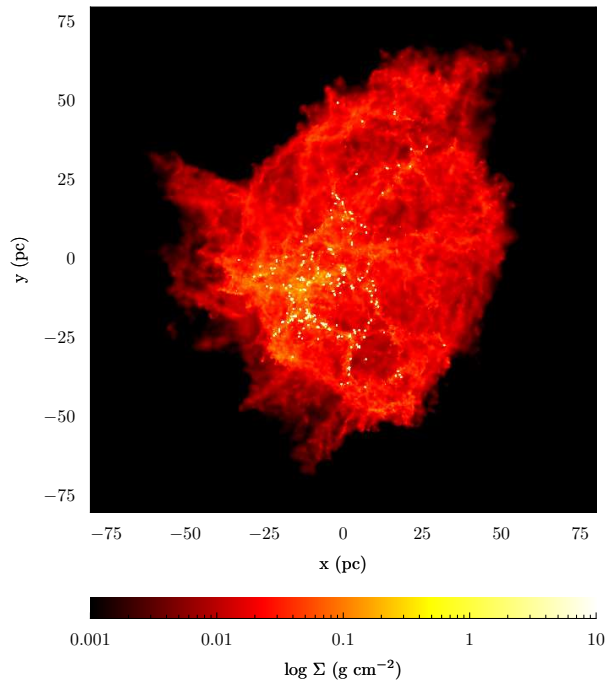}
\caption{Column--density map viewed down the $z$--axis of the system at the point where ionizing radiation was turned on. \label{fig:init_snap}}
\end{figure}
\begin{figure}
\includegraphics[width=0.45\textwidth]{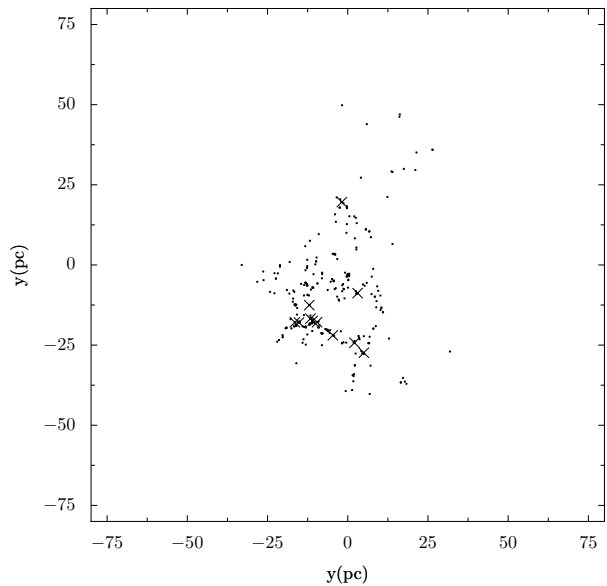}
\caption{Positions of clusters (those massive enough to be ionizing sources are marked as crosses) at the point when ionization was turned on.\label{fig:sources}}
\end{figure}
\begin{figure*}
     \centering
     \subfigure[Column density plots of cold neutral gas and stars (left panel), cold neutral gas and hot ionized gas (centre panel) and hot ionized gas alone (left panel) 0.66 Myr after ionization was turned on.]{\includegraphics[width=\textwidth]{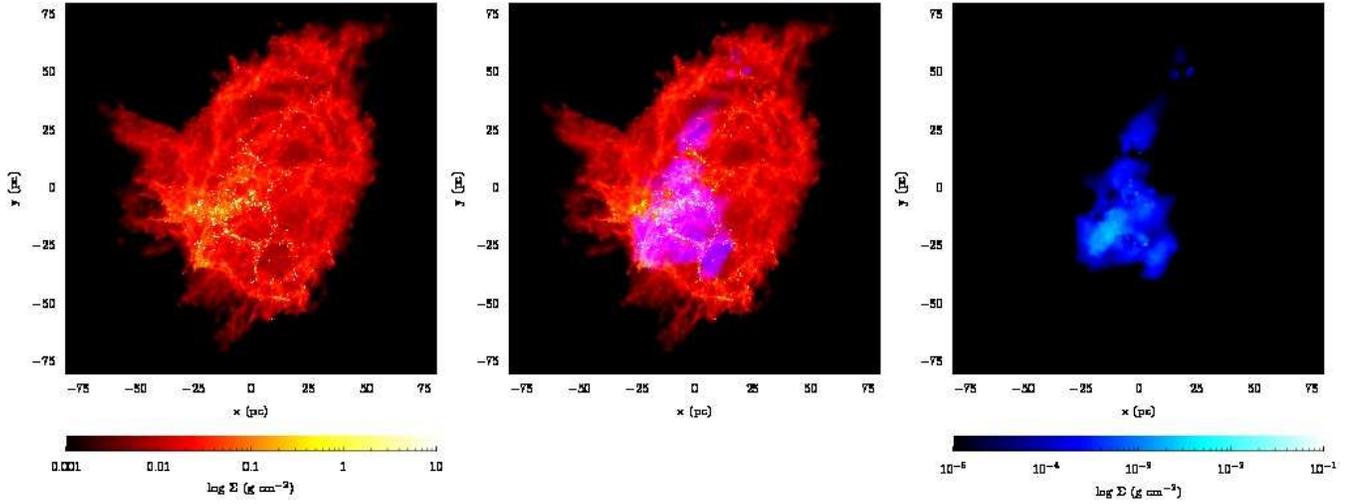}}
     \vspace{.1in}
     \subfigure[Column density plots of cold neutral gas and stars (left panel), cold neutral gas and hot ionized gas (centre panel) and hot ionized gas alone (left panel) 1.08 Myr after ionization was turned on.]{\includegraphics[width=\textwidth]{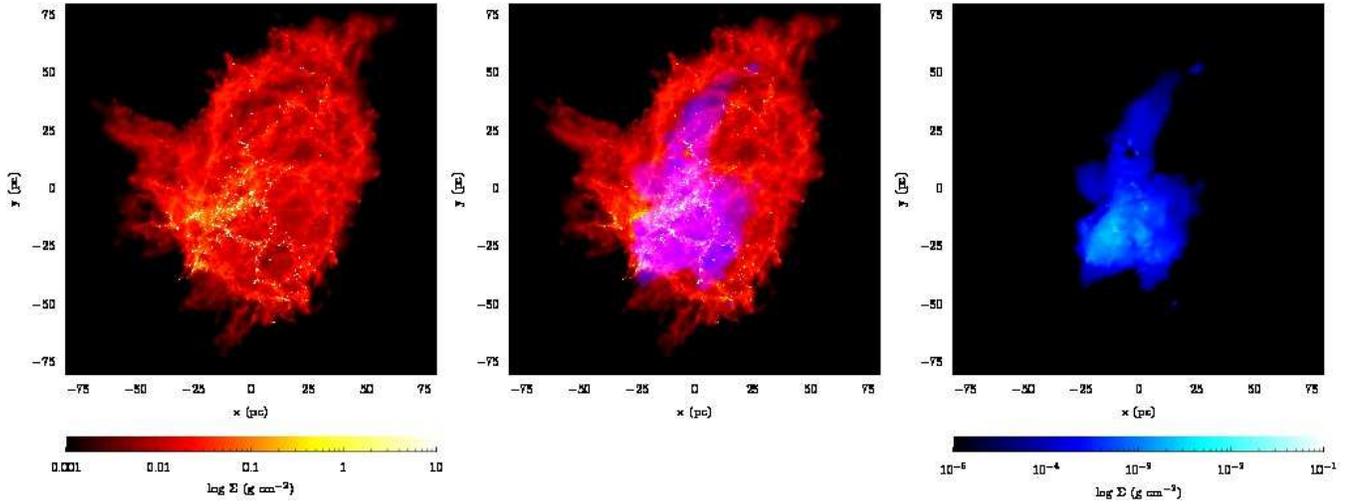}}
     \vspace{.1in}
     \subfigure[Column density plots of cold neutral gas and stars (left panel), cold neutral gas and hot ionized gas (centre panel) and hot ionized gas alone (left panel) 2.18 Myr after ionization was turned on.]{\includegraphics[width=\textwidth]{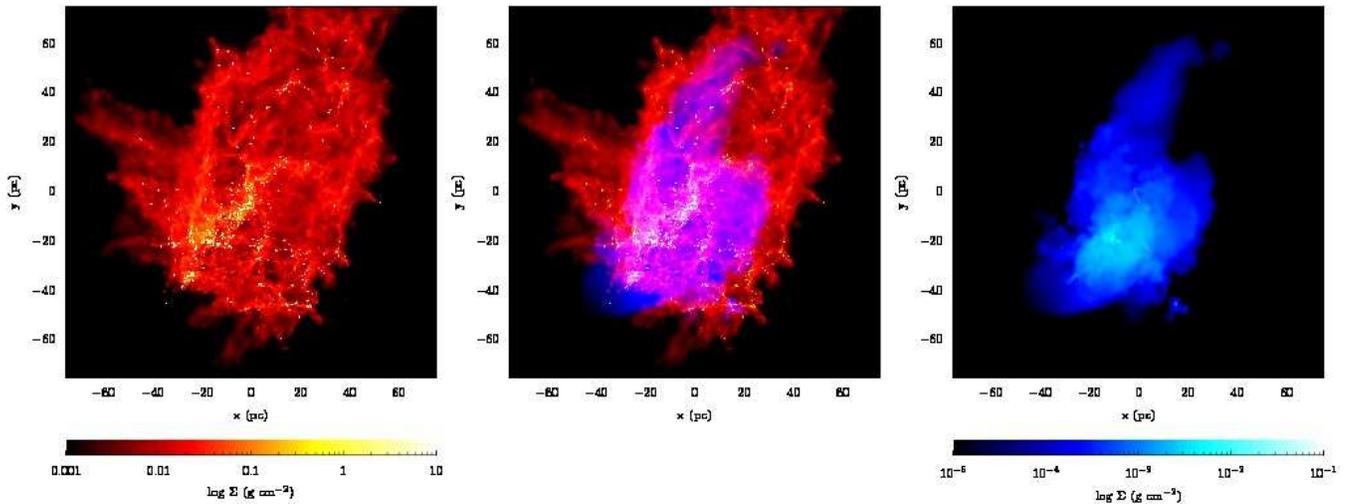}}
     \caption{Evolution of the ionized and neutral gas. \label{fig:HII_plots}}
\end{figure*}

\indent We perform a separate low--resolution calculation in which we enable ionization when the system has an age of 3.7 Myr (0.64 t$_{\rm ff}$), which is approximately the time when the system begins to give rise to clusters large enough to host O--stars. At this point in the system's evolution, there are eleven such clusters and these therefore form our ionizing sources. Figure \ref{fig:init_snap} shows a column--density plot as viewed down the $z$--axis of the system at this epoch, with white dots representing clusters. The gas distribution is composed of a network of dense filaments in which most of the star formation is occurring, particularly at their junctions, permeated by a froth of low--density voids. In Figure \ref{fig:sources}, plotted to the same linear dimensions, we show the positions of the ionizing sources (obviously as time progresses and clusters accrete, more objects acquire enough mass to become ionizing sources).\\
\indent In Figure \ref{fig:HII_plots}, we show at three different epochs column density maps of the cold gas (red in the left and centre panels ) and the HII (blue in the right and centre panels). It is clear that the hot, low--density HII fills up cavities in the cold gas, so that the HII bubbles are illuminated from the edge rather than the centre, often by several widely--separated O--stars, and that several small HII regions eventually join to form a single large and irregularly--shaped one. The brightest part of the HII region is not centred on the main concentration of clusters, but instead abuts it. Additionally, as shown in Figure \ref{fig:init_snap}, these cavities existed before the ionizing sources were turned on and were generated by the turbulent velocity field with which the cloud was seeded, so they have not been created by the action of feedback. We also note that the boundaries of the cavities are delineated by the dense filaments in which star formation is taking place, so that the morphology of the system appears to consist of several overlapping HII--filled bubbles with vigorous star formation activity on their borders. Several authors \citep[e.g.][]{2008ApJ...688.1142K,2009A&A...503..107P} have taken this configuration to be indicative of triggered star formation but in our calculations, this is clearly not so. The overall HII morphology is irregular, with an approximately round core centred at about (-10,-10) pc and approximately 25 pc across, and several spurs and offshoots, one extending about 50 pc towards the top right in the final frame of the simulation. The distribution of HII is also markedly clumpy due to the clumpiness of the gas and the action of multiple ionizing sources. In the final frame of Figure \ref{fig:HII_plots}, it is clear that radiation and ionized gas are leaking out of the cluster through holes in the neutral material, particularly in the bottom left of Figure 6c. This structure superficially resembles a champagne flow, but its origin is clearly rather different, since the bubble was not created by the HII region.\\
\indent We find, as in \cite{2005MNRAS.358..291D}, that the effect of the ionization is strongly constrained by the dense filamentary gas in which all the ionizing sources are embedded, and on which it has only a small effect. Ionization fronts propagate only into the low--density gas, so that only a small amount of dense material is ionized, but most of the HII is generated from lower--density gas. In Figure \ref{fig:dens_compare}, we plot the final density versus the initial density for all ionized gas (red) and for a randomly--chosen one percent of the neutral gas. In general, the density of the neutral gas has increased during the simulation, since the cloud is collapsing and forming stars. Conversely, the density of ionized gas has generally decreased (although not in all cases, as some material has become denser than it was initially \textit{before} being ionized). In Figure \ref{fig:cumass_ions}, we plot the cumulative mass of ionized gas against its initial density. The ionized gas is clearly drawn from material that was of a low density initially and much of it has subsequently become less dense as it expands into the voids between the filaments.\\
\indent The reason that photoionization has such a limited effect on the dense gas in the cluster is that all the ionizing sources are embedded in, and continuously accreting from, dense material and most of their photon fluxes are absorbed by the accretion flows, leading to evolution very different from the classical picture of a champagne flow. \cite{1995RMxAC...1..137W} gave a simple formula for a critical accretion rate $\dot{M}_{\rm crit}$ onto an ionizing source of mass $M$ and ionizing luminosity $L$ (Lyman continuum photons per second) above which all the ionizing photons are absorbed by the accretion flow, so that the HII region cannot expand and may instead contract:
\begin{eqnarray}
\dot{M}_{\rm crit}>\left(\frac{4\pi LGMm^{2}_{\rm H}}{\alpha_{\rm B}}\right)^{\frac{1}{2}}.
\end{eqnarray}
\begin{figure*}
     \centering
     \subfigure[]{\includegraphics[width=0.31\textwidth]{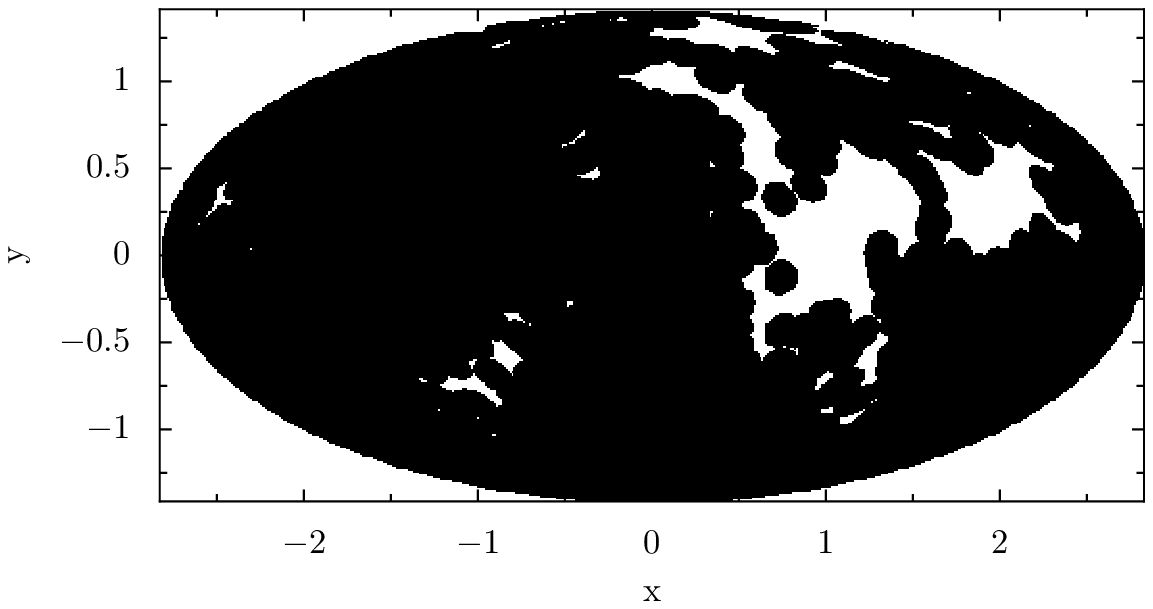}}
     \hspace{.1in}
     \subfigure[]{\includegraphics[width=0.31\textwidth]{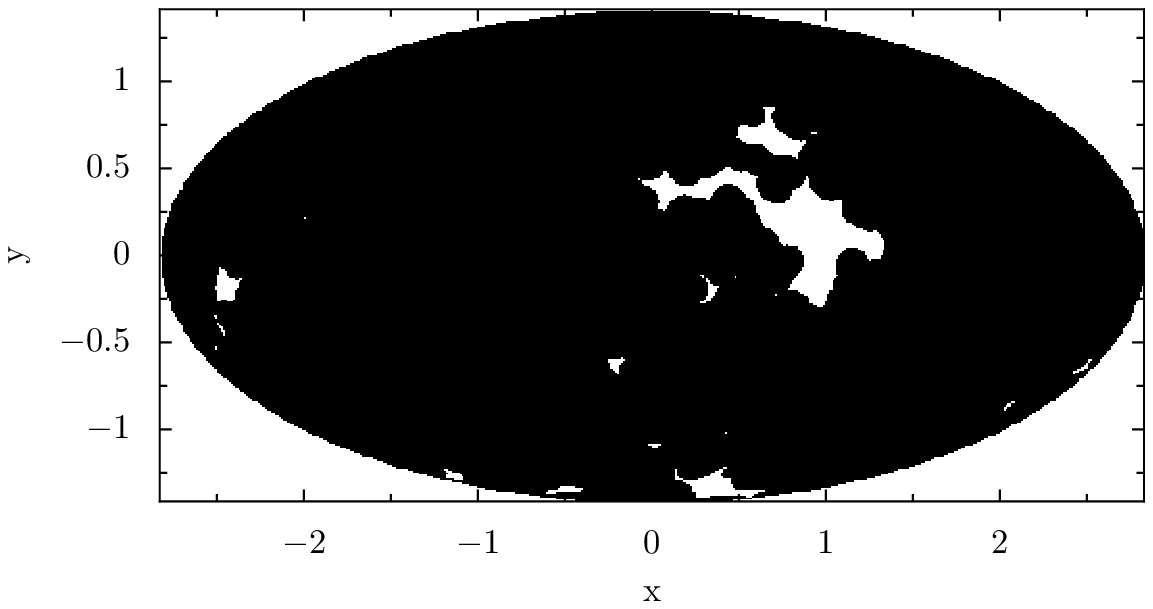}}
     \hspace{.1in}
     \subfigure[]{\includegraphics[width=0.31\textwidth]{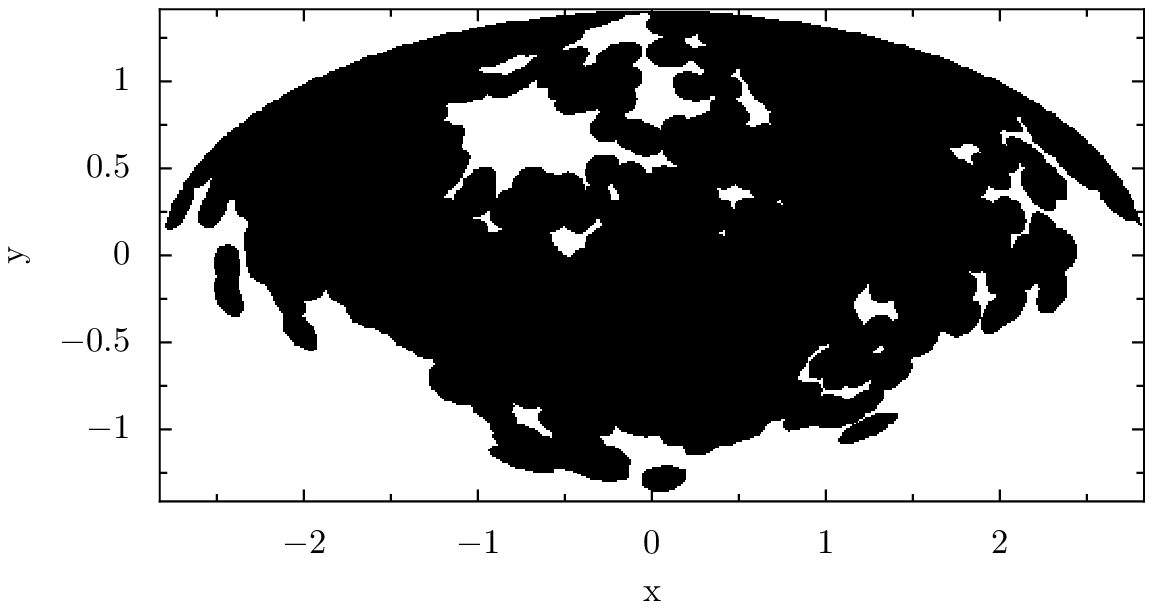}}
\caption{Hammer projections showing directions in which ionizing radiation is absorbed before reaching a radius of 5pc (black areas) from the point of view of three sources at the time when ionizing radiation was switched on.\label{fig:aniso1}}
\end{figure*}     
\begin{figure*}
     \centering
     \subfigure[]{\includegraphics[width=0.31\textwidth]{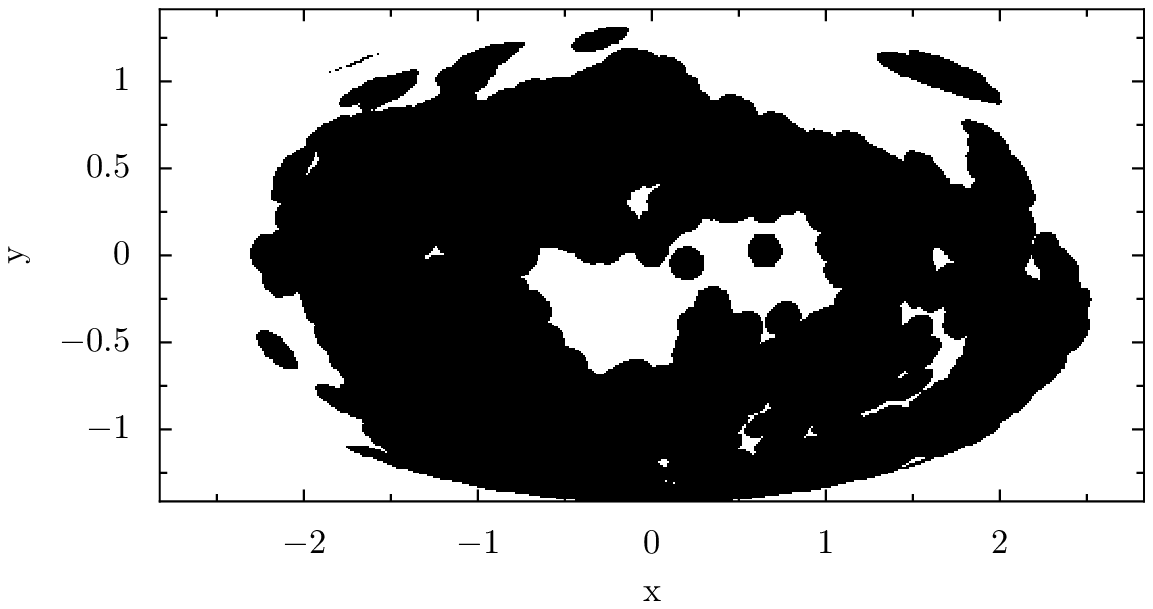}}
     \hspace{.1in}
     \subfigure[]{\includegraphics[width=0.31\textwidth]{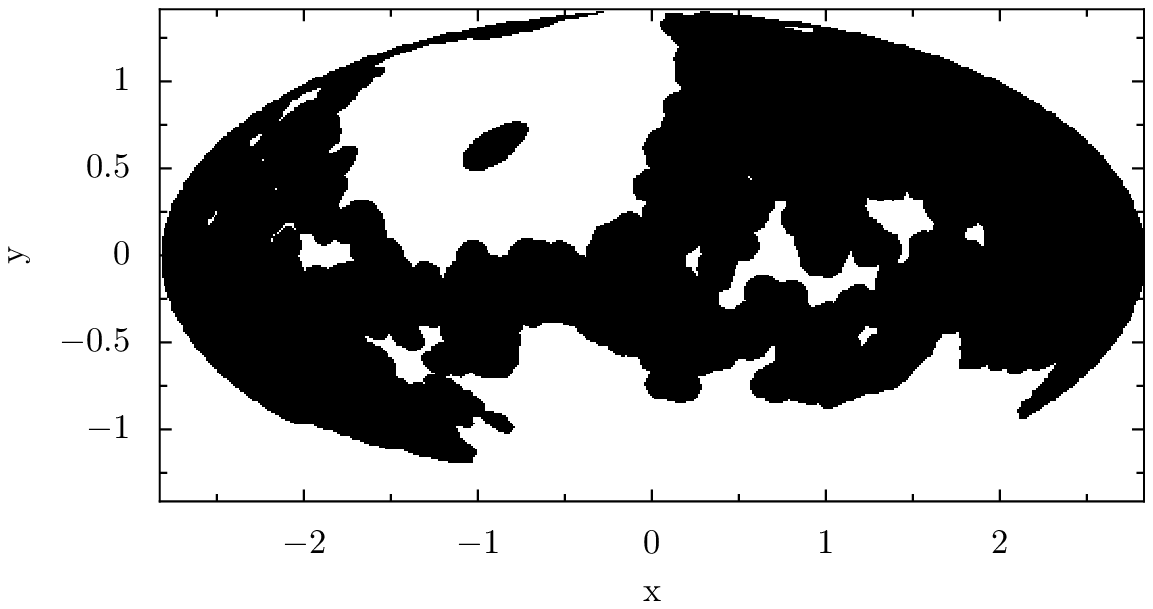}}
     \hspace{.1in}
     \subfigure[]{\includegraphics[width=0.31\textwidth]{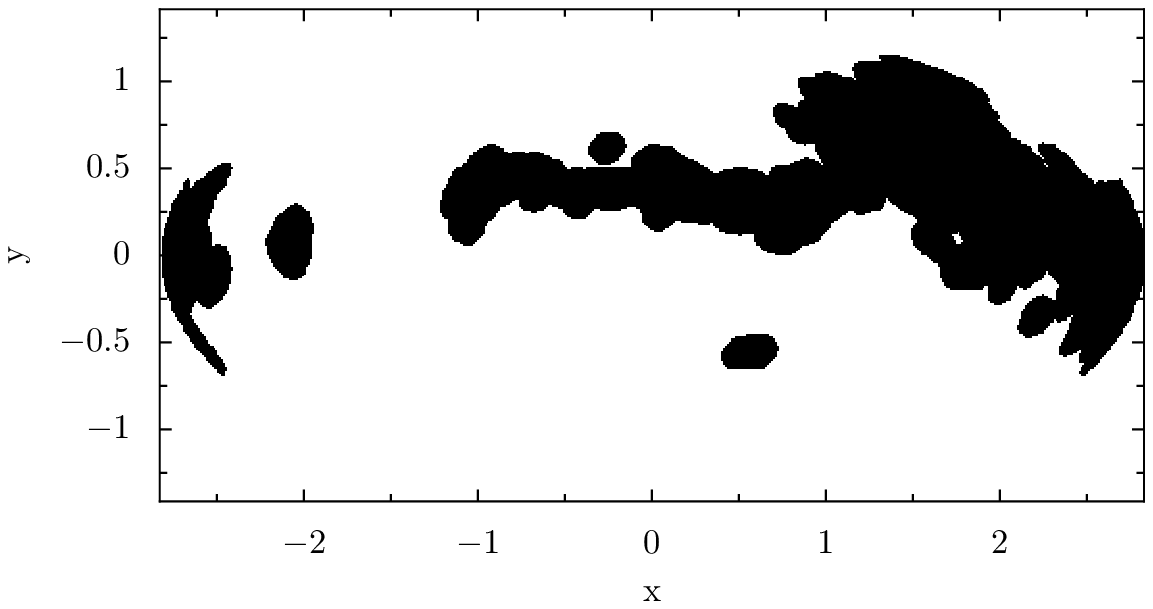}}
\caption{Hammer projections showing directions in which ionizing radiation is absorbed before reaching a radius of 5pc (black areas) from the point of view of three sources 1.62 Myr after ionizing radiation was switched on. \label{fig:aniso2}}
\end{figure*}       
We find that for all our sources, at virtually all times, their accretion rates are supercritical. If the accretion flows were spherically symmetric, they would then absorb all of the combined photon flux of the sources in the cluster. There are three ways in which such an accretion--swamped ionizing source may still grow an HII region. The first and most obvious is that the accretion rate may decline. Although the accretion flows in our simulation are by no means steady, the accretion rates do not decrease with time. The second possibility is that, as the source accretes and grows in mass, its ionizing luminosity increases as a function of the mass. The ionizing flux may then eventually overwhelm the accretion of neutral material, and this does play a role in our simulation. The third way in which a swamped source may still ionize some of its surroundings occurs if the accretion flow is anisotropic, so that radiation may escape from the source in some directions. It is this third way that is of most importance here; the dense gas around our ionizing sources is highly anisotropic. In Figures 7 and 8, we plot at two different epochs (at the time ionization was switched on, and 1.62 Myr later), Hammer projections depicting, from the point of view of randomly--chosen ionization sources, the directions in which ionizing radiation is able to penetrate the surrounding gas to a depth further than 5pc. Areas coloured black are those in which the radiation is absorbed before reaching this radius. For all the sources, large fractions of the sky are opaque to their radiation. At the earlier epoch, the average fraction of sky opaque beyond 5 pc as seen by all sources is $\sim70\%$. At the later epoch, this fraction has dropped to $\sim45\%$, so clearly the effect of ionization increases as time passes, but rather slowly.\\
\indent Since the accretion flows are too vigorous to be disrupted by the ionizing radiation and they cover a large fraction of the sky as seen form most of the ionizing sources, the effect of feedback on most of the gas in the cluster, and hence on the star formation process is small. We terminated the simulation after ionization has been active for $\approx2.2$Myr, at which point the age of the system is $\approx 6$Myr and the first supernova to occur is due in $\approx0.8$Myr, at which point the first O--stars will reach 3Myr in age and leave the main sequence. By this time, $\sim2\times10^{4}$M$_{\odot}$ of gas was unbound, about two of thirds of it ionized, representing only $\sim2\%$ of the system's total mass. This corresponds to a rate of mass loss of $\sim10^{-2}$M$_{\odot}$yr$^{-1}$. Given that the cluster contains on average 10s to a few hundred O--stars, this rate of mass--loss is rather low compared to those reported by, e.g., \cite{1980A&A....90...65M}, due to the quenching effect on the feedback of the accretion flows onto the ionizing sources. Although we have only allowed ionization to act for $\sim2.2$ Myr, we think it highly unlikely that the effects of ionization will change significantly in the remaining time until supernovae begin to explode in the protocluster.\\
\section{Discussion and Conclusions}
We find that in large stellar systems whose formation has been triggered by the decay of a turbulent velocity field, star formation produces a complex hierarchical structure with clustering on many different length and mass scales. Star formation generally follows the dense filaments generated by the velocity field, with the most massive and densest clusters found at the junctions of the filaments. Conversely, the protocluster exhibits large voids in which no star formation occurs at all. The freefall time of our model system is $\sim4$Myr, so that dynamical interactions are unable to erase the large--scale structure of the protocluster on the timescale on which stellar feedback is expected to act and in particular cannot greatly influence such structure before the first supernovae begin to detonate. The configuration of stellar clusters and gas produced by the turbulent velocity field therefore defines the backdrop against which feedback operates.\\
\begin{figure}
\includegraphics[width=0.45\textwidth]{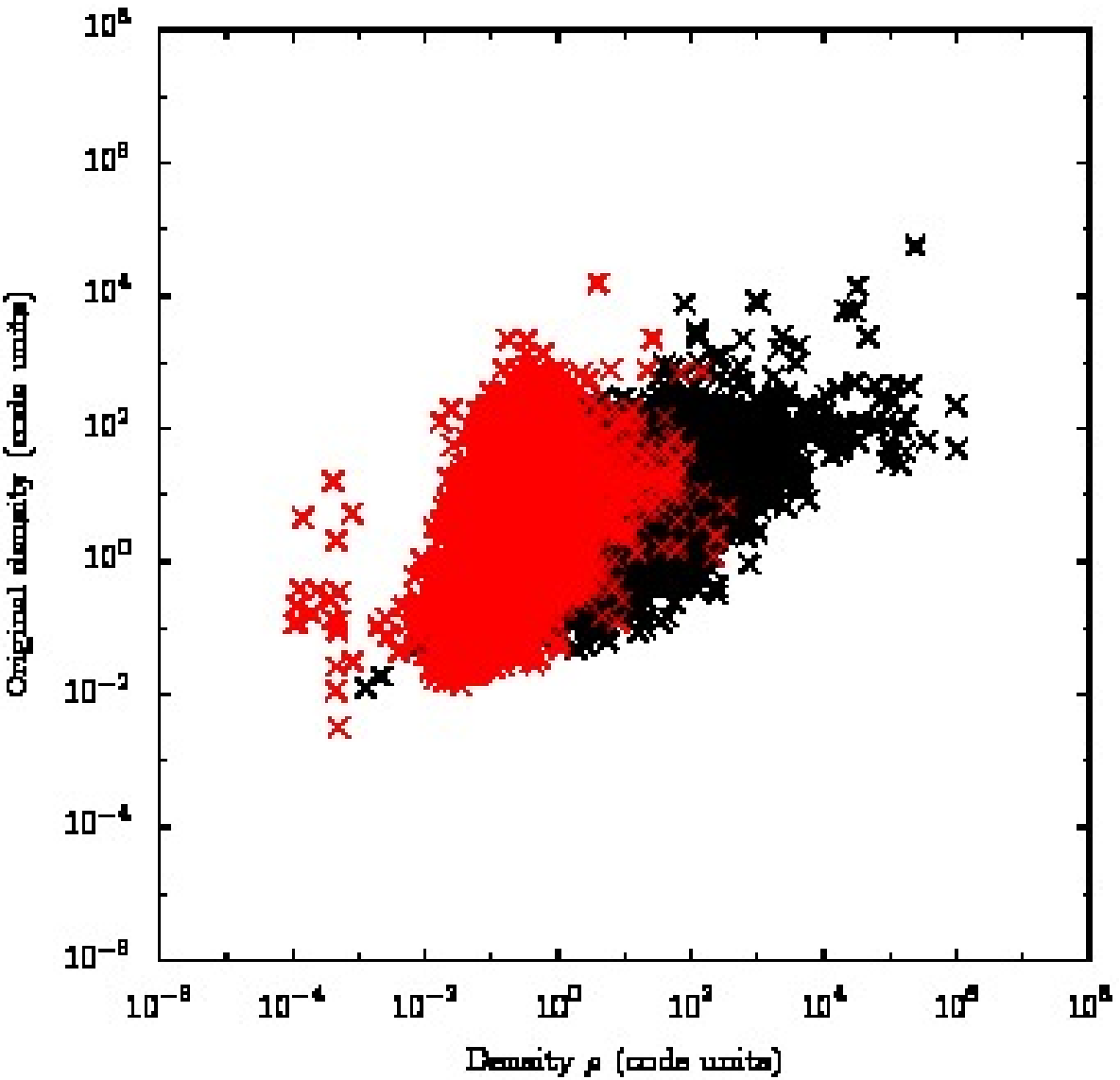}
\caption{Comparison of the initial ($y$--axis) and final ($x$--axis) densities of randomly--selected neutral (black) and ionized (red) particles.\label{fig:dens_compare}}
\end{figure}
\begin{figure}
\includegraphics[width=0.45\textwidth]{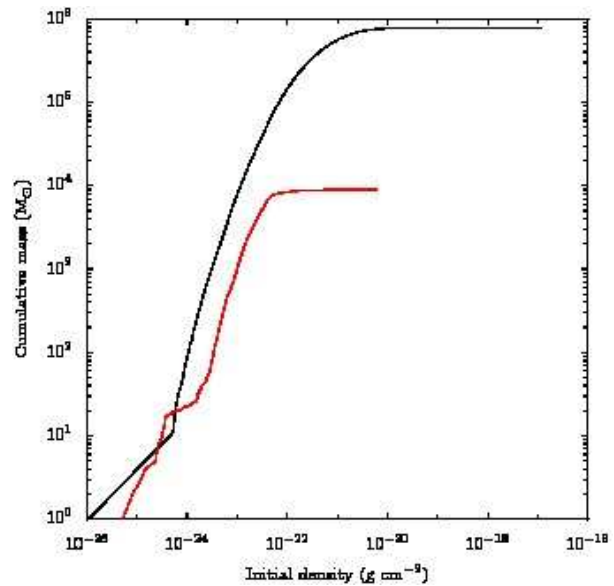}
\caption{Plot of cumulative mass against initial density for the neutral (black) and ionized (red) gas.\label{fig:cumass_ions}}
\end{figure}
\begin{figure}
\includegraphics[width=0.45\textwidth]{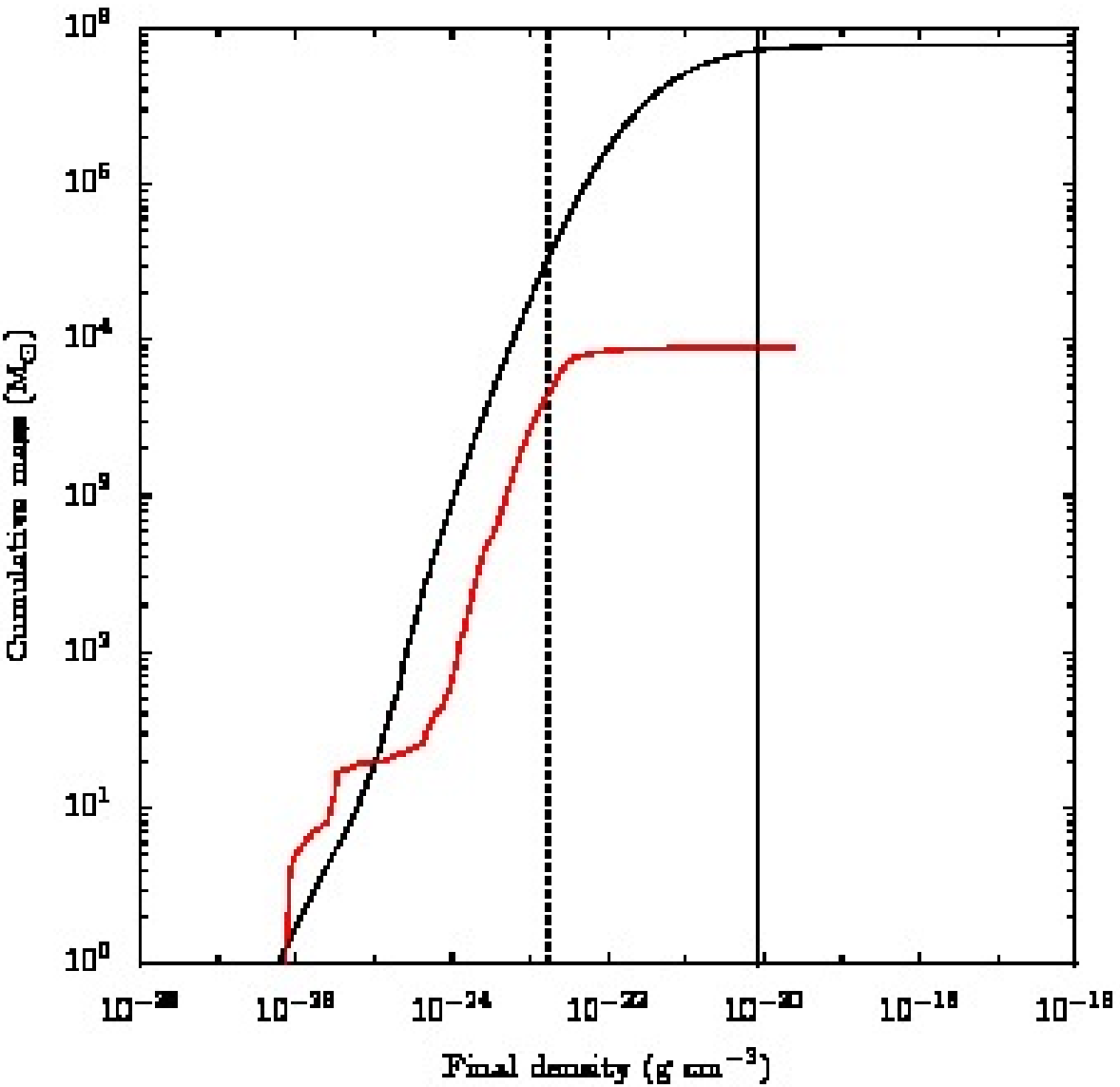}
\caption{Cumulative mass plotted against density for randomly--selected neutral (black) and ionized (red) particles 1.62 Myr after ionization was turned on. The dashed vertical line indicates the median density of the ionized gas and the vertical dotted line indicates the densest cold gas whose pressure would be less than that of the median--density ionized gas if the cold gas were all at 40K. Neutral gas to the right of the dotted line cannot therefore be strongly influenced by pressure from ionized gas.\label{fig:cum_rho}}
\end{figure}
\indent Photoionizing radiation has little impact on the evolution of our model protocluster and in particular is unable to disrupt the dense gas in which most of the star formation is occurring due to the presence of strong accretion flows onto the ionizing sources. As a result,  photoionizing feedback on this system is restricted to filling pre--existing voids in the cold gas with hot ionized material and the principal dynamical effect of feedback derives from the pressure exerted by this hot gas on the neutral gas. We compute at several times the average density of the ionized gas, which would have a temperature of $10^{4}$K and select all the neutral particles whose thermal pressure is in excess of the average thermal pressure of the ionized gas. The particles so selected are then too dense to be affected by the pressure of ionized gas. We find in this way that only a small amount, $\approx3\%$ of the gas in the cluster at any time is too dense to be affected by the low--density ionized gas produced by the O--stars. We illustrate this qualitatively in Figure \ref{fig:cum_rho} where we plot the cumulative mass against density for neutral gas in black and ionize gas in red 1.62Myr after ionization was turned on. The vertical dashed line is the median density of the ionized gas and the vertical dotted line is the densest neutral gas whose pressure is less than that in ionized gas at the median density, assuming all neutral gas has a temperature of 40K. Neutral gas to the right of the dotted line would then be too dense to be strongly affected by the thermal pressure in the ionized gas.\\
\indent In principle then, the ionization should be able to significantly influence the evolution of the protocluster's gas reserves. In Figure \ref{fig:toodense}, we plot the positions of the gas whose density is too large to be affected by the HII. We neglect here the ram--pressure of the cold gas against which the HII must also contend, since the gas is infalling towards the ionizing sources. The dense gas clearly (and not surprisingly) correlates strongly with the positions of the clusters in the simulation -- the stars are to be found embedded in the densest gas. In order to significantly affect the evolution of the cluster, photoionization would need to clear this material away. Since it can do so only very slowly, as shown in Figures \ref{fig:aniso1} and \ref{fig:aniso2}, the hot gas instead leaks into the low--density voids.\\
\begin{figure}
\includegraphics[width=0.45\textwidth]{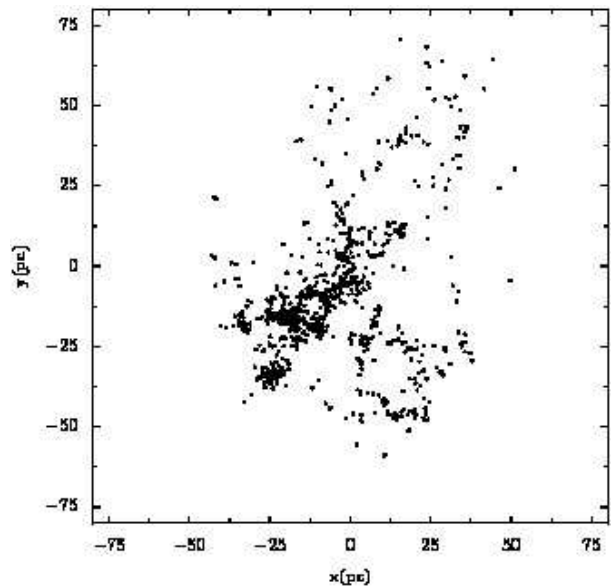}
\caption{Positions of the $\approx3\%$ of the gas too dense to be influenced by the thermal pressure of ionized gas at the time when ionization is turned on.\label{fig:toodense}}
\end{figure}
\indent Classical models \citep[e.g][]{1979ApJ...233...85B,1987sbge.proc..467L,2002ApJ...566..302M} indicate that ionization--driven champagne flows should be an efficient mechanism for disrupting molecular clouds and this is very likely to be true in many systems, such as Orion \citep{2005ApJ...627..813H}, where O--stars are located near the edges of clouds. However, in cases where O--stars are embedded in highly inhomogeneous and often filamentary gas deep in cloud interiors, photoionization may not be so effective. In this work and that of several other authors \citep[e.g][]{1989A&A...216..207Y, 2005MNRAS.358..291D,2010ApJ...711.1017P,2010ApJ...719..831P}, embedded O--stars struggle to disrupt their natal clusters by photoionization. This does not imply that this must always be the case, but it does suggest that gas expulsion by ionizing radiation is more difficult than implied by earlier models. This difference emerges principally because strong accretion flows onto the ionizing sources suppress the growth of HII regions, a process which was previously often neglected.\\
\indent Stellar winds may assist the UV radiation \citep[e.g][]{1997A&A...326.1195C,2001PASP..113..677C,2003ApJ...594..888F}, although we found in \cite{2008MNRAS.391....2D} that cluster disruption by winds alone may be a rather slow process. The oldest O--stars in our simulation will soon explode as supernovae and it is possible that the mechanical and energetic feedback from the few hundred O--stars that reside in our clusters would succeed where photoionization has failed. \cite{1985A&A...145...70T}, for example, concluded that a single supernova could disrupt $\sim10^{4}$ of molecular cloud material. The gravitational binding energy of our model system is a few$\times10^{51}$ erg and it has formed the equivalent of a few hundred O--stars by an age of 6Myr, so that supernovae should in principle have no trouble unbinding it. Alternatively, as suggested and investigated by \cite{2005MNRAS.359..809C}, molecular clouds and their embedded clusters (or cluster complexes) may not be bound in the first place.
\section{Acknowledgements}
JED acknowledges support from a Marie Curie fellowship as part of the European Commission FP6 Research Training Network `Constellation' under contract MRTN--CT--2006--035890, and from the Institutional Research Plan AV0Z10030501 of the Academy of Sciences of the Czech Republic and project LC06014--Centre for Theoretical Astrophysics of the Ministry of Education, Youth and Sports of the Czech Republic.

\bibliography{myrefs}

\label{lastpage}

\end{document}